\documentclass[12pt]{article}
\usepackage{epsfig,epsf}
\begin{document}
\def\beq{\begin{equation}}
\def\eeq{\end{equation}}
\def\eq#1{{Eq.~(\ref{#1})}}
\def\fig#1{{Fig.~\ref{#1}}}
\newcommand{\as}{\alpha_S}
\newcommand{\bra}[1]{\langle #1 |}
\newcommand{\ket}[1]{|#1\rangle}
\newcommand{\bracket}[2]{\langle #1|#2\rangle}
\newcommand{\intp}[1]{\int \frac{d^4 #1}{(2\pi)^4}}
\newcommand{\mn}{{\mu\nu}}
\newcommand{\tr}{{\rm tr}}
\newcommand{\Tr}{{\rm Tr}}
\newcommand{\T} {\mbox{T}}
\newcommand{\braket}[2]{\langle #1|#2\rangle}
\newcommand{\ab}{\bar{\alpha}_S}

\begin{flushright}
BNL-NT-02/22\\
October 23, 2002\\
\end{flushright}

\begin{center}
{ {\Large \bf  Parton saturation and $\mathbf{N_{part}}$ scaling \\
of semi--hard processes in QCD}
\vskip1cm
Dmitri Kharzeev${}^{a)}$,  Eugene Levin${}^{b),c)}$ and Larry
McLerran${}^{a)}$}
\vskip1cm

{\it a) Nuclear Theory Group,}\\
{\it Physics Department,}\\
{\it Brookhaven National Laboratory,}\\
{\it Upton, NY 11973 - 5000, USA}\\
\vskip0.3cm
{\it b ) HEP Department, School of Physics,}\\
{\it Raymond and Beverly Sackler Faculty of Exact Science,}\\
{\it Tel Aviv University, Tel Aviv 69978, Israel}\\
\vskip0.3cm
{\it ${}^{c)}$ DESY Theory Group}\\
{\it 22603, Hamburg, Germany}\

\end{center}
\bigskip
\begin{abstract}
We argue that the suppression of high $p_t$ hadrons discovered recently in heavy ion collisions 
at RHIC may be a consequence of saturation in the Color Glass Condensate.
We qualitatively and semi-quantitatively describe the data, in particular, the
dependence upon the number of nucleon participants. 
We show that if parton saturation sets in at sufficiently small energy, then in 
nucleus--nucleus collisions at RHIC and LHC energies the cross sections 
of semi--hard processes should scale approximately with the number of participants, $N_{part}$. 
Our results provide a possible explanation of both the absence of apparent jet quenching at 
SPS energies and its presence at RHIC.  
Under the same assumption we predict that in semi--central and central $pA$ ($dA$) collisions 
at collider energies the dependence of semi--hard processes on the number of 
participating nucleons of the nucleus will change to $\sim (N_{part}^A)^{1/2}$.   
The forthcoming data on $dA$ collisions will provide a crucial test of this description.
\end{abstract}
\vskip0.3cm

The recent results from RHIC \cite{brahms,phenix,phobos,star} suggest
that relativistic heavy ion
collisions  at high energies probe QCD in the  non--linear regime of high
parton density and strong  color fields
\cite{GLR,MUQI,BM,MV}.
Parton dynamics in this new regime is quite different from what is expected in
perturbative QCD, and is best described by
{\it parton saturation} \cite{GLR,MUQI,MV} resulting in a formation of a new type of high density
matter,  the  Colour Glass Condensate (CGC)
\cite{MV,YM1,YM2,YM3,YM4,KV,KN}.

In this high density regime,
the transition amplitudes are dominated not by quantum fluctuations, but by
the configurations of classical field containing large, $\sim 1/\alpha_s$,
numbers of gluons.   Even though the coupling $\alpha_s$ becomes
small due to the high density of partons, the fields interact strongly due to the
classical coherence.
One thus uncovers new
non-linear features of QCD,
which cannot be investigated in a more traditional perturbative approach.

This new dynamics appears to be consistent \cite{KN,KL,KLN} with the  RHIC data on
the multiplicity distributions of
produced hadrons as a function of centrality, rapidity, and collision energy.
In the Color Glass Condensate picture, the approximate scaling of
hadron multiplicity with the number of participants $N_{part}$,
and the logarithmic deviation from it, $\sim \ln N_{part}$ stemming from the running
of the QCD coupling as a function of parton density, has been interpreted as a consequence
of parton saturation \cite{KN,KL,KLN}. In
this letter we will show that QCD saturation gives a very substantial contribution to
 the suppression of the single particle inclusive
 distributions even at quite large transverse momenta,
and can be responsible for (at least a part of) the observed approximate $N_{part}$ scaling of processes
at high $p_t$.

\vskip0.3cm

The experiments on high $p_t$ hadron production show \cite{phenixpt,starpt,phobospt} that the inclusive
cross section is strongly suppressed with respect to the scaling with the number of binary nucleon--nucleon
collisions $N_{coll}$ expected on the
basis of the factorization theorem \cite{FT} for hard processes
in perturbative QCD (pQCD). An illuminating way to study this effect is to look at
the dependence of hadron yields in a given $p_t$ bin as a function of centrality.
The data seem to be closer to the approximate scaling with the
number of participants, previously established for small $p_t$ particles dominating
the total multiplicity.
It is indeed a surprise that this
behavior characteristic of soft  processes persists up to the transverse momenta
($p_t \sim 5 \div 10$ GeV), where we expect the fragmentation of the incoherently produced jets
to dominate.

The approximate $N_{part}$ scaling
is of course a restatement of the observed earlier suppression of the high
$p_t$ single particle inclusive distributions. One possible explanation of the observed
suppression is the predicted quenching of the produced jets in hot quark--gluon matter \cite{Bj,GW,BDMPS}.
The observation
of approximate $N_{part}$ scaling in the entire range of hadron momenta up to $\sim 6$ GeV
is however problematic for the scenario in which the jet
energy loss is the only reason behind the suppression. Indeed,
let us assume, following the pQCD approach, that most of the hadrons produced at RHIC originate
from mini-jet fragmentation. If mini-jet production is a hard incoherent process,
one expects the initial number of mini-jets to scale with the number of binary collisions $N_{coll}$.
The energy loss in QCD is dominated by the induced gluon radiation and therefore as the produced
mini-jets lose energy they will emit additional softer gluons.
These emitted gluons in turn will fragment into hadrons, increasing further the total multiplicity
of hadrons produced in the event. If the initial number of mini-jets were proportional to $N_{coll}$,
the measured total multiplicity would then grow even faster with centrality, in a marked
contradiction to the experimentally observed behavior. This contradiction can be avoided only
if we assume that the {\it{initial}} production of moderate $p_t$ mini-jets, the fragmentation of which
gives the dominant contribution to the measured multiplicity, is suppressed with respect to
$N_{coll}$ scaling. (The jets with really high $p_t$ because of the small production cross section
do not contribute to the total hadron multiplicity in any significant way, and their production
according to $N_{coll}$ scaling, and subsequent energy loss, would not lead to any conflict with
the observed scaling of the total multiplicity).

\vskip0.3cm

The main goal of
our letter is to show that the approximate $N_{part}$ scaling of mini-jet production at moderately
high $p_t$ (up to $p_t \sim 6 \div 8$ GeV at RHIC energies) has a natural
explanation in terms of parton saturation and the Color Glass Condensate. We also
demonstrate
that the
$N_{part}$ scaling exhibits very non--trivial properties of dense partonic systems not only
at momentum transfers around the saturation scale $Q_s$, but also at much higher momenta, naively as
large as $\sim Q_s^2/\Lambda_{QCD}$.
  For heavy nuclei, this new momentum scale is ${\cal{O}}(5-10\ {\rm GeV})$ at RHIC energies, and
even larger at the energies of the LHC,  ${\cal{O}}(25-50\ {\rm GeV})$.  We will in fact find
that the correct scale for nuclei is somewhat reduced relative to this estimate, but is still
large compared to $Q_{sat}$.  We find this scale to be about $4~$GeV at RHIC and of order
$10~$ GeV at the LHC.

Our arguments are based on the following foundations.
First, in the Color Glass Condensate, the saturated
parton density provides a new dimensionful scale; as a consequence, the structure functions
at momentum transfer $Q^2$ depend on a single variable $\tau \equiv Q_s^2/Q^2$ (``geometrical scaling'').
This has been proven \cite{GLR,MUQI,MV,BL,SGK,LT1,ILM,KT} for $Q < Q_s$. Moreover, it has been
observed that the deep inelastic scattering data at small $x \leq 0.01$ exhibit geometrical scaling in the
{\it entire} kinematical region of $Q^2$ accessible at HERA, and it has been shown  \cite{IIM}
that the scaling indeed holds in a wide range of $Q_s^2 \leq Q^2 \leq Q_s^4/\Lambda_{QCD}^2$.
Second, the properties of QCD evolution are modified in the domain of high parton density, where
geometrical scaling holds. As we will discuss below, the anomalous dimension of the gluon
density in this region is approximately equal to $1/2$, as was computed in \cite{GLR,BL,LT1,IIM}.
Third, in the range of  $Q_s^2 \leq Q^2 \leq Q_s^4/\Lambda_{QCD}^2$ the saturation scale for
nuclear targets scales according to $Q_s^2 \sim A^{1/3}$, a behavior which was argued to be responsible
for the centrality dependence of hadron multiplicity at RHIC \cite{KN,KL}.

\vskip0.3cm

Let us begin by considering an external probe with virtuality $Q^2$ interacting with a target
of longitudinal size $L$ and an average (3-dimensional) parton density $n$. The probability
of interaction in the target is determined by the ratio $\kappa \equiv L/\lambda$ of the target size
to the mean free path of the probe $\lambda = (\sigma n)^{-1}$,
where $\sigma \sim \alpha_s(Q^2)/Q^2$ is the cross section of the probe scattering off a parton.
 At small Bjorken $x$, when the coherence length
$\sim 1/(mx)$ ($m$ is the nucleon mass) exceeds the target size,
the probe can interact coherently with partons
located at different longitudinal coordinates,
and the only relevant parameter characterizing the target
becomes the density of partons in the
transverse plane, $\rho = n L$;
for a nuclear target, this implies $\rho \sim A^{1/3}$. The interaction probability
is then determined by a dimensionless ``parton packing factor''
$\kappa \sim   \rho(x) \ \alpha_s(Q^2) / Q^2$. The boundary of the saturation region, and the
corresponding value of ``saturation scale'' $Q_s^2(x)$,  are
given by the line in the $(x, Q^2)$ plane along which the interaction probability is of order one,
$\kappa(x, Q^2) \simeq 1$.

We will now write the packing factor in the form \cite{KN} suited for nucleus--nucleus collisions, where a
centrality cut can be used to select events with different transverse densities $\rho_{part}$  of nucleons participating
in the collision (``participants''):
\beq \label{PF}
\kappa \,\,=\,\,\frac{8\,\pi^2\,N_c}{N^2_c - 1}
\frac{\alpha_S(Q^2)}{Q^2}
xG(x,Q^2) \,\frac{\rho_{part}}{2}
\eeq
where $xG(x,Q^2_s)$ is the gluon structure function; we assume that $x$ is sufficiently small
for gluons to dominate the parton densities. Consider the kinematical regime of small $x$ and high $Q^2$
in which both
$\alpha_s \ \ln Q^2$ and $\alpha_s\ \ln (1/x)$ are large; in this domain both logarithms have to be
resummed, and
the solution to DGLAP evolution equations  \cite{DGLAP}  
for $xG$ can be written in the ``double log
approximation'' (DLA), which allows us to re-write
\eq{PF} in the following naive form, neglecting the influence of saturation effects on the QCD evolution:
\beq \label{DLAPF}
\kappa \,\,=\,\,\exp\left[\sqrt{4 \bar{\alpha}_S \,( y - y_0)\,\xi}
\,\,-\,\,\xi
\,\,+\,\,\xi_{part}\right];
\eeq
we have assumed that at the initial point $x_0$ of the evolution in
$x$ the transverse momentum squared of partons is $Q^2_0$, and
have introduced the following notations:
$y$ is rapidity, with $y \,-\,y_0\,\,=\,\,\ln(x_0/x)$, $\xi\,=\,\ln(Q^2/Q^2_0)$,
$\xi_{part}
\,=\,\ln
(\rho_{part}/\rho_0)\,\,\approx\,\,\frac{1}{3} \ln A$, $A$ is the atomic
number, $\rho_0$ is parton density in a nucleon,
and $\bar{\alpha}_S \, = \,(N_c/\pi)\as$ in the large $N_c$ approximation.

We now investigate the effect of saturation on the function $\kappa$.
Our treatment so far assumed, in the spirit of pQCD,
that evolution starts at some initial fixed scale $Q_0^2 \sim \Lambda_{QCD}^2$
separating perturbative and non--perturbative domains, and that $Q^2 \gg Q_0^2$. Traditionally, one assumes
that the dynamics at the scale $Q_0^2$ and below is driven by non--perturbative phenomena and should
be described by {\it{a priori}}
unspecified initial conditions for the structure functions, which are then constrained
by experimental data. The parton saturation approach is different in that it provides theoretical
tools to {\it{compute}} the parton distributions once their density becomes large enough. If saturation occurs
in the nuclear wave function at some $x_0$ and at the corresponding scale $Q_s^2(x_0)$, in performing QCD
evolution of parton densities to smaller values of $x$ and larger values of $Q^2$ we should therefore
provide the saturated Color Glass Condensate distribution as the initial condition.

\vskip0.3cm

It is convenient to recall the well--known property of DGLAP equation  \cite{DGLAP}  (see
Ref.\cite{GLR} for example) which allows to write
its solutions in a form expressing their Green's function--like properties:
\beq \label{TOTINT}
\kappa(y,\xi) \,\,=\,\,\int^{\xi}\,\,d \xi' \ \kappa(y -
y_0, \xi - \xi')\,\,\kappa(y_0,\xi')
 \eeq
Using the DLA form of the solution we see that the integral in (\ref{TOTINT})
has a saddle point at $\xi' = \xi (y_0/y)$. If we consider $y \gg y_0$ ($x \ll x_0$), then the
dominant contribution to the integral comes thus from $\xi' \to 0$, corresponding to the initial
condition at $Q_0^2 \ll Q^2$. If, however, $x_0$ is sufficiently small for the saturation to set in,
the initial condition will be provided by McLerran-Venugopalan model \cite{MV} at the scale
 $\xi' = \xi_{sat}(y_0) = \ln(Q_s^2(y_0)/Q_0^2)$. The
integral over $\xi'$ will then be dominated by $\xi'\,\, \approx\,\, \xi_{sat}(y_0)$.
This results in a different solution, which instead of  \eq{DLAPF} is now given by
\beq \label{DLANEW}
\kappa \,\,=\,\,\exp \left[\sqrt{4 \bar{\alpha}_S \,( y - y_0)\,(\xi
\,-\,\xi_{sat}(y_0))}
\,\,-\,\,\xi
\,\,+\,\,\xi_{sat}(y_0)\right]
\eeq

The saturation boundary can be found from the condition
$\kappa  \,=\, 1$.
%
This leads to the following equation for the saturation scale at rapidity $y$:
\beq \label{SATSCNEW}
 4\,\,\bar{\alpha}_S\,(\,\xi_{sat}(y) \,-\,\xi_{sat}(y_0)\,)\,(y -
y_0)\,=\,(\xi_{sat}(y) \,-\,\xi_{sat}(y_0))^2\,\,.
\eeq
This equation has a simple solution
\beq \label{SCSOL}
\xi_{sat}(y)\,\,=\,\,\,4\,\,\bar{\alpha}_S\,\,(y\,-\,y_0)
\,\,+\,\,\xi_{sat}(y_0)\,\,.
\eeq
Since $Q^2_s(x_0)$ is proportional to $A^{\frac{1}{3}}$ at relatively low
energies, when
$y \ \approx \ y_0$
(see \eq{SATSCNEW}
or the McLerran -- Venugopalan formula \cite{MV})), one can still
conclude that $Q^2_s(y)\,\,\propto\,\,A^{\frac{1}{3}}$. For nuclear collisions, this means that
the product of the nuclear overlap area $S_A$ and the saturation scale is proportional to the
number of participants,
$S_A\,Q^2_s \,\,\propto\,\,N_{part}$ \cite{KN,KL}.

\vskip0.3cm

Let us establish the leading behavior of the solution (\ref{DLANEW}) at high $Q^2$.
At very high $\xi \gg 2 \xi_{sat}(y)$, the solution tends to 
\beq \label{CANON}
\kappa \sim {Q_0^2 \over Q^2},
\eeq
modified by
the double logarithmic factor of $\exp\left(\sqrt{4\overline \alpha_S(y-y_0)\xi}\right)$.
This result does not exhibit geometrical scaling, and is the same as the naive perturbative result.
Consider, however, the region where
\beq
\xi - \xi_s(y) \ll \xi_s(y) - \xi_s(y_0),
\eeq
which corresponds to a vast kinematical domain of
\beq \label{DOM}
Q^2 \ll {Q_s^4 (y) \over Q_s^2 (y_0)};
\eeq
note that this is precisely the condition for the observed geometrical scaling to hold in
deep--inelastic scattering \cite{IIM}.
In this domain, we should expand the exponent of Eq.(\ref{DLANEW}) in powers of the ratio
$\Delta \xi / (\xi_{sat} (y) - \xi_{sat} (y_0))\ \ll \ 1$, where we have introduced the notation
$\Delta \xi = \xi -\xi_{sat} (y) =  \,\ln(Q^2/Q^2_s)$. The expansion to first order
yields
$$
\sqrt{4 \bar{\alpha}_S \,( y - y_0)\,(\xi
\,-\,\xi_{sat}(y_0))}
\,\,-\,\,\xi
\,\,+\,\,\xi_{sat}(y_0) =
$$
\beq \label{DLAEXP}
= \left( \xi_{sat}(y) - \xi_{sat}(y_0) \right)\ \sqrt{1 + {\Delta \xi \over
\xi_{sat}(y) - \xi_{sat}(y_0)}} - \Delta \xi - ( \xi_{sat}(y) - \xi_{sat}(y_0)) \simeq
- {1 \over 2}\ \Delta \xi,
\eeq
where we have used Eq.(\ref{SCSOL}).
This result translates into the following $Q^2$ dependence of the packing factor:
\beq \label{ANOM}
        \kappa = \left({{Q_{sat}^2(y)} \over {Q^2}} \right)^{1 \over 2}.
\eeq
The appearance of the square root clearly shows that the properties of QCD evolution are dramatically
affected by the saturation phenomena in the entire region (\ref{DOM}), much more broad than
the originally anticipated  $Q^2 \sim Q_s^2(y)$. The comparison of (\ref{CANON}) and (\ref{ANOM})
reveals that in the presence of Color Glass Condensate the  gluon
density acquires the anomalous dimension equal to $1/2$.

\vskip0.3cm

\begin{figure}
\begin{minipage}{10cm}
\begin{center}
\epsfxsize=8cm
\epsfysize=6.5cm
\hbox{ \epsffile{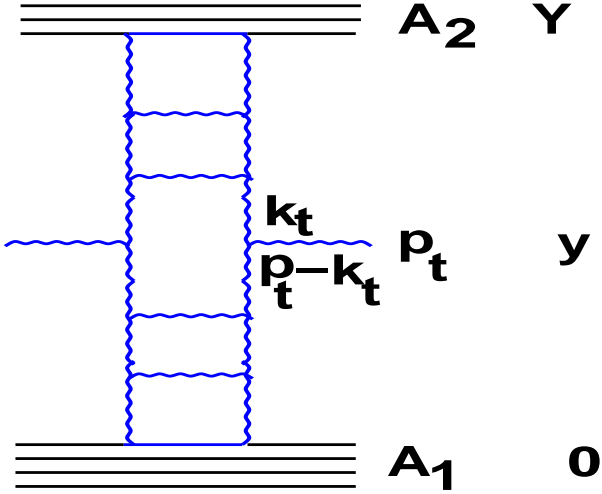}}
\end{center}
\end{minipage}
\begin{minipage}{5.5cm}
\caption{The simplest Mueller diagram for the high $p_t$ jet
inclusive
production. }
 \label{mud}
\end{minipage}
\end{figure}

Let us now turn to the discussion of semi--hard processes in hadron and nuclear collisions.
 Due to the AGK cancellation \cite{AGK}, the inclusive cross section
for large $p_t$ jet production is described by the Mueller diagram shown
in \fig{mud}. It can be written as (see
Refs.\cite{GLR,LR1,LL,KR,GM,KM,BRAUN,K00,KT} )
\beq \label{GENX}
E {d \sigma \over d^3 p} = {4 \pi N_c \over N_c^2 - 1}\ {1 \over p_t^2}\
 \int d k_t^2 \
\alpha_S \, \varphi_{A_2}(Y - y, k_t^2)\, \varphi_{A_1}(y,
(p-k)_t^2),
\eeq
where $\varphi$'s are the unintegrated parton distributions related to the packing factor $\kappa$
discussed above by simple relation $\varphi \sim S_A \kappa_A / \alpha_s$, and
$Y$ is the rapidity of the measured jet.
Using the formula Eq.(\ref{ANOM}) for  $\varphi \sim S_A \kappa_A / \alpha_s$
in the domain (\ref{DOM}), and noting that in this case $Q^2 = p_t^2$,
we see that \eq{GENX} leads to
\beq \label{NPARTSC}
\frac{d N}{d y d^2 p_t}\,\, = \ \frac{1}{S_A}E {d \sigma \over d^3 p} \,\
\propto\,\,S_A Q^2_s/p^2_t\,\,\rightarrow\,\,N_{part}/p^2_t
\eeq
where we have used that $N_{part} \propto Q^2_s \,S_A$.
Note that the number of produced hard particles scales with the number
of participants.
The $N_{part}$ scaling given by \eq{NPARTSC} can be formulated in the following way:
$$
\frac{1}{N_{part}}\,\frac{d N}{d y d^2 p_t}\,\,\,\,\,\,\,\,\,\,\,\,
 \ \mbox{\rm is almost independent of centrality cut for }  
\,\,\,\,\,\,\frac{Q^2_s(y)}{\Lambda}\,\,\geq \,\,p_t\,\,\geq
\,\,Q_s(y)\,\,,
$$
where $\Lambda = Q_s(y_0)$ is the saturation scale at some $y_0 < y$ marking the onset of saturation.
  This is the central result of our paper.


Of course, ``genuine'' hard processes still scale with the number of collisions. Indeed,
consider $p_t > Q_s^2 / \Lambda$; then according to Eq. (\ref{CANON}),
 $\varphi \,\,\propto\,\,Q^2_s/p^2_t$ and
the formula \eq{GENX} yields the number of hard particles proportional to
 $S_A Q^4_s/p^4_t
\,\sim \,N_{coll}/p^4_t$ in accord with the pQCD factorization theorem.

\vskip0.3cm

We have demonstrated that at high energies corresponding to $y \gg y_0$ the effects of the
Color Glass Condensate persist up to very large values of $p_t$;
how far do these effects extend at RHIC and LHC energies? The answer depends on the energy at which
the saturation effects set in. It seems reasonable to assume, on the basis of the analysis of
centrality and energy dependence of hadron multiplicities \cite{KN,KL,KLN}, that the minimal
value of saturation scale is about $Q_s^2(y_0) \simeq 0.5\ {\rm GeV}^2$. In central $Au-Au$ collisions
at RHIC energy, where $Q_s^2(y) \simeq 2\ {\rm GeV}^2$, the boundary of the kinematical domain fully dominated
 by saturation extends thus up to $p_t \simeq \sqrt{2}\ Q_s^2(y) / Q_s(y_0) \simeq 4$ GeV.
At LHC, where $Q_s^2(y) \simeq 4.5\ {\rm GeV}^2$ \cite{KLN}, this boundary extends up to $p_t \simeq 9$ GeV.
It is also important to note that in peripheral collisions, as $Q_s^2(y)$ becomes small and
approaches the value of $Q_s^2(y_0)$, we return to the familiar situation in which the boundary
of hard processes sets in at a rather small value of $p_t \sim 1$ GeV, and the number of produced
mini-jets, and hadrons, thus scales proportionally to the number of binary collisions.
However, as the centrality increases, and $Q_s^2(y)$ grows, the number of the produced
semi-hard particles begins to deviate from the binary collision scaling; the transition to the
$N_{part}$ scaling occurs.

\vskip0.3cm

While these arguments are well defined and grounded theoretically,
reliable computations will require further improvement
of the quantitative understanding of saturated parton distributions. Nevertheless, we will
present a calculation of the semi-hard particle production using a simple input distribution stemming
from \eq{DLANEW}:
\beq \label{DIST1}
\varphi_A (y, p_t^2) = \exp \left[\sqrt{4 \bar{\alpha}_S \,( y - y_0)\,\xi(p_t^2)}\
\right]\ \ \varphi^0_A (y_0, p_t^2);
\eeq
where $\xi(p_t^2) = \ln(p_t^2/Q_s^2(y_0))$. 
To bring this formula in line with the standard description of the deep-inelastic scattering (DIS) data, 
we have to modify it by slightly changing the anomalous dimension. Indeed, it has been 
demonstrated \cite{EKL} that the following simple form of the anomalous dimension 
in the Mellin moment variable $\omega$ (in the 
leading order in $\alpha_s$): 
\begin{equation}
\gamma(\omega)\,\,=\,\,\alpha_S\,\left(\,\frac{1}{\omega} - 1\right) \label{phenan}
\end{equation}
describes the DIS data well. 
This ``phenomenological" anomalous dimension leads to a multiplicative factor of $(Q_s^2/Q^2)^{\alpha_S}$ 
in our formula in the $x$-$p_t$ representation. 
The resulting 
anomalous dimension is consistent with the result found from analysis of
the BFKL equation and saturation in leading and next to leading order \cite{EKL,MU02,BFKL,NLOBFKL,CCS}.
We use the running coupling in (\ref{phenan}), freezing its value at small virtualities 
at $\alpha_s^{max} = 0.45$; therefore, at very high energies, when $\alpha_s < \alpha_s^{max}$ in the entire region (\ref{DOM}), 
the additional multiplicative factor turns into a constant, and we return to the anomalous 
dimension of $1/2$.

For the function $\varphi^0_A (y_0, p_t^2)$, describing
 parton distributions in the Color Glass Condensate at the onset of saturation, we will use the
formula
$$
\varphi^0_A(x,p^2_t) =  \frac{S_A}{\as}
\,\,d(\tau\,=\,\frac{p^2_t}{Q^2_s(x)})\,\,;
$$
\beq \label{DISF}
d(\tau)  =  \frac{(2\,\tau \,+\,1)}{\sqrt{4\,\tau\,+\,1}}\,\ln\frac{
\sqrt{4\,\tau\,+\,1} \,+\,1}{\sqrt{4\,\tau\,+\,1}
\,-\,1}\,\,-\,\,1\,\,.
\eeq
where $S_A$ is the nuclear overlap area. At large $p_t$, this distribution falls off as $1/p_t^2$,
corresponding to the classical bremsstrahlung, and at small $p_t$ it has the behavior $\sim \ln(Q_s^2/p_t^2)$,
established previously for the Color Glass Condensate \cite{LT1,ILM}.
It agrees well with the numerical solution of non--linear evolution
equations \cite{LT1}; the derivation of \eq{DISF}
and detailed discussion of its properties will be presented elsewhere
\cite{KDIS}. The value of the initial rapidity $y_0$ should be chosen to correspond to the onset
of coherent partonic phenomena; we choose $y_0 = y_b - \ln (1/x_0)$ ($y_b$ is the beam momentum)
corresponding to the value of Bjorken $x_0 \simeq 0.1$; this marks the beginning of the regime in which
the coherence length $\sim 1/(m x_0)$ starts to exceed the typical inter-nucleon distance.

\begin{figure}
\begin{center}
\epsfxsize=16.5cm
\epsfysize=12.5cm
\hbox{ \epsffile{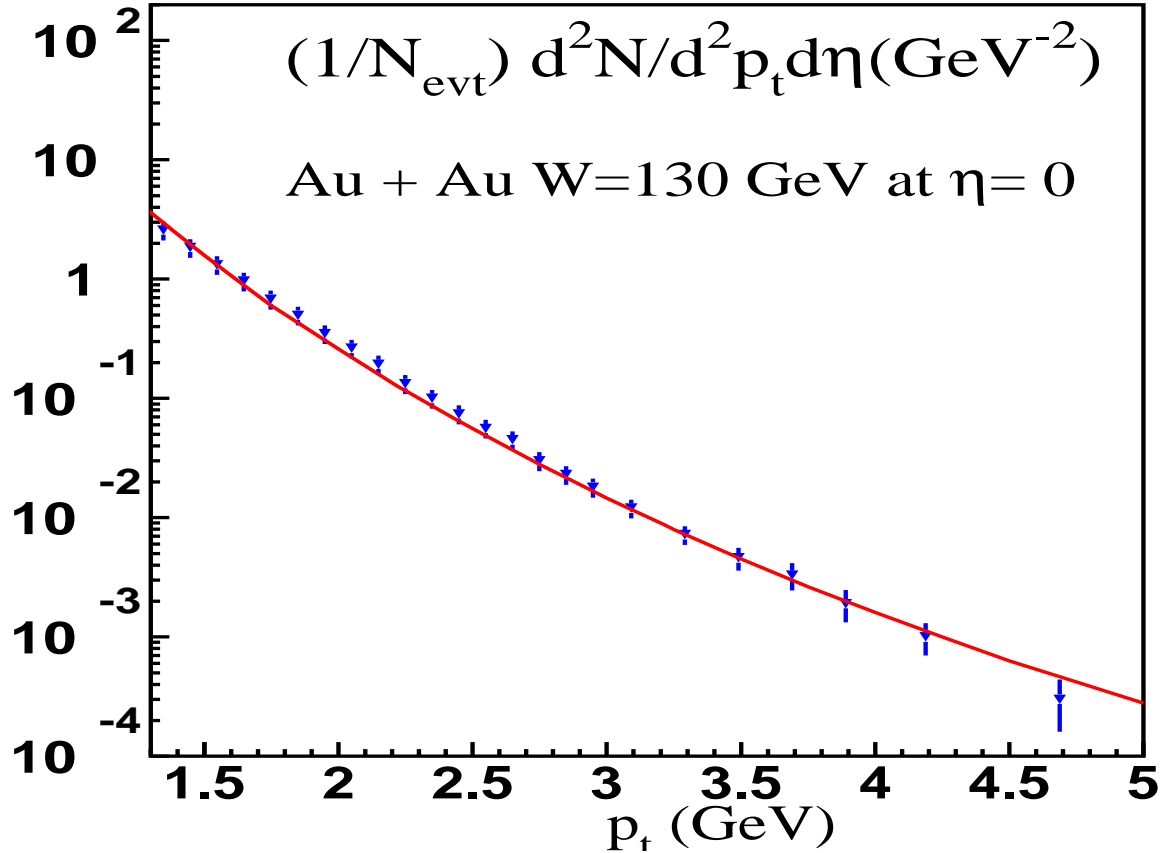}}
\end{center}
\caption{The $p_t$-spectrum of charged hadrons in central ($0-10\%$ centrality cut) $Au-Au$ collisions 
at $\sqrt{s} = 130$ GeV. The data is from \cite{phenixpt}.}
 \label{pt}
\end{figure}

\begin{figure}
\begin{center}
\epsfxsize=16.5cm
\epsfysize=16.5cm
\hbox{ \epsffile{ 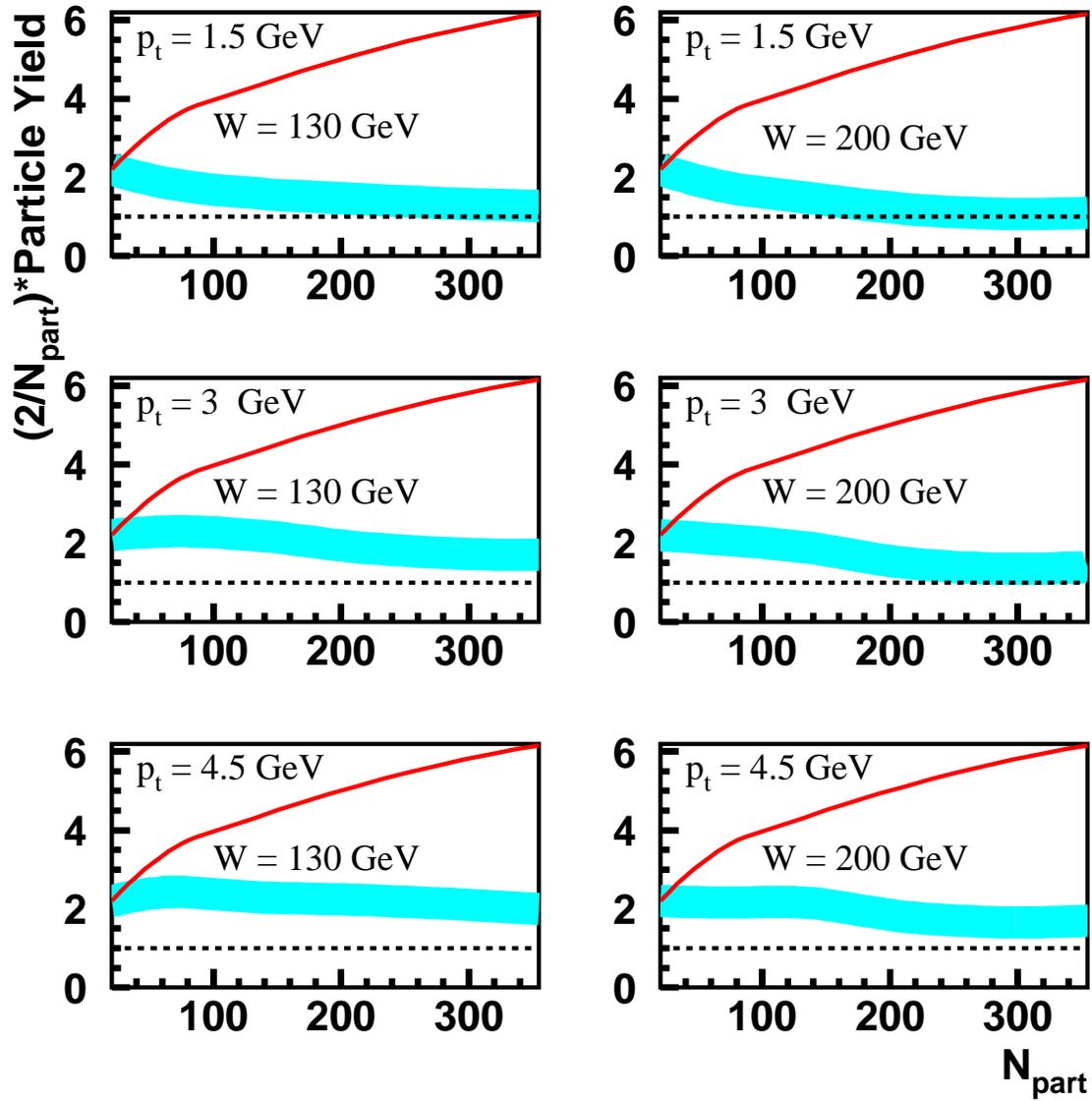}}
\end{center}
\caption{Centrality dependence of hadron yields per participant pair in $Au-Au$ collisions 
at $\sqrt{s}=130$ and $200$ GeV in 
different $p_t$ bins; the yields are normalized to the yield in peripheral collisions. Upper solid lines 
show the behavior expected in perturbative QCD.}
 \label{cent}
\end{figure}

\begin{figure}
\begin{center}
\epsfxsize=16.5cm
\epsfysize=16.5cm
\hbox{ \epsffile{ 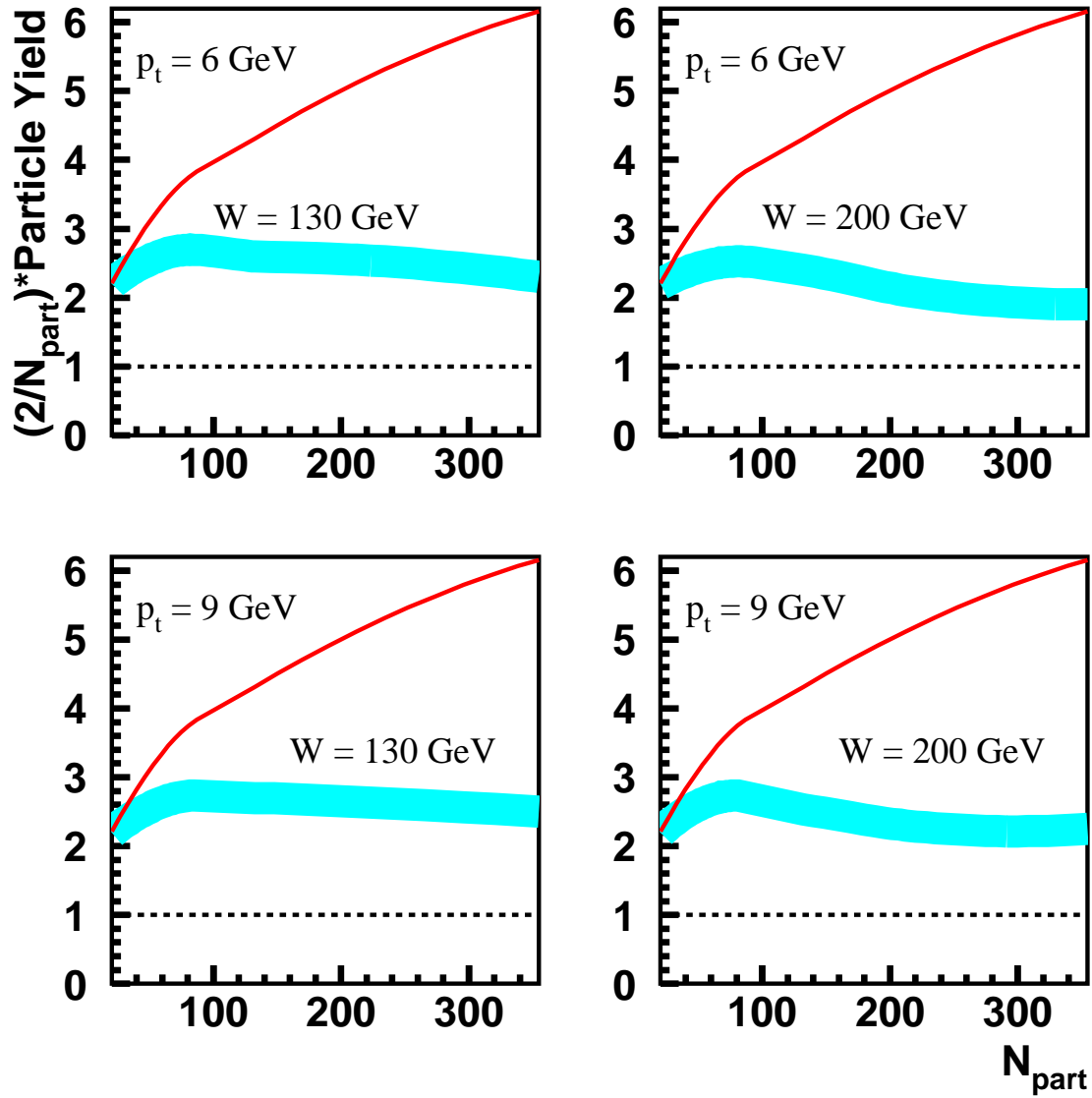}}
\end{center}
\caption{Same as in Fig.\ref{cent} in the $p_t$ bins of $6$ and $9$ GeV.}
 \label{cent1}
\end{figure}

The cross section of hadron production can then be  obtained by convoluting the result of calculation according
to Eqs.(\ref{GENX},\ref{DIST1}) with the jet fragmentation function $D_{frag}(z, p_t)$:
\beq \label{HADPR}
\frac{d \sigma^{hadron}}{d y d p^2_t}\,\,\,=\,\,\,\int \, dz \,\frac{d
\sigma^{jet}}{d y d q^2_t}\,\,\delta(
\,p^2_t\,\,-\,\,z^2\,\,q^2_t\,)\,\,D_{frag}(z, p_t)\,\,;
\eeq
we also introduce in the distribution functions the common $\sim (1-x)^4$ factors enforcing
the behavior at $x \to 1$ prescribed by the quark counting rules.

\vskip0.3cm

The result for the charged hadron spectrum in central $Au-Au$ collisions at $\sqrt{s} = 130$ GeV is
compared in Fig.\ref{pt} to the experimental data from PHENIX Collaboration.
While the agreement is not perfect, the calculated
cross section does agree with the data within error bars (only statistical error bars are shown),
which is remarkable in spite of the crude
approximations we made and our use of the standard fragmentation functions \cite{KKP}
fitted for the use with pQCD calculations. In Fig. \ref{cent}, we show also the dependence
of particle yields in different $p_t$ bins on centrality. While the result admittedly depends
on the approximations we made and is not very robust, it does show  that up to
quite large values of $p_t$ the yield of particles deviates strongly from the scaling with binary
collisions, shown by the upper curve on Fig.\ref{cent}.

Further work is clearly needed before a reliable calculation of high $p_t$ cross sections
in the Color Glass Condensate approach can be
performed. Nevertheless, it is already clear that high density QCD effects have to be taken
into account in the interpretation of the discovered at RHIC suppression of high $p_t$
particles.

\vskip0.3cm
The mechanism proposed here may also explain the old puzzle \cite{Wang:1998hs}: 
the absence of apparent jet quenching in the SPS data 
\cite{Albrecht:1998yc}, \cite{Aggarwal:1998vh} on high $p_t$ neutral pion production. 
Indeed, at relatively small energy of the SPS ($\sqrt{s} \simeq 17$ GeV) the transverse momentum 
spectra of the produced particles are rather steep, and even a small jet energy loss would induce 
large deviations from the perturbative scaling with the number of collisions. These deviations were 
not observed: 
the data appear consistent with the perturbative QCD calculations \cite{Wang:1998hs}. 
At the same time, many of the ``soft'' properties of the collisions, as well as the suppression of 
charmonium, point to the presence of strong 
collective effects (see, e.g., \cite{Heinz:2000bk}). 

Basing on the fact that the prediction of KLN model \cite{KLN} for centrality dependence of hadron multiplicities in 
the $\sqrt{s}=20$ GeV 
run at RHIC appeared to be successful \cite{Baker:2002ft}, it may be possible to assume that saturation 
sets in around SPS energy\footnote{We do not believe that extrapolation to still lower energies would 
make sense since the coherence length becomes much smaller than nuclear size in that case.}. Since saturation 
provides favorable initial conditions for thermalization \cite{Baier:2000sb},\cite{Baier:2002bt}, 
the emergence of collective effects in ``soft'' observables and charmonium suppression would thus be 
likely. However, since for $y \to y_0$ the domain (\ref{DOM}) $Q_s^2(y) < Q^2 < Q_s^4(y)/Q_s^2(y_0)$ 
shrinks to zero, the hard processes above the saturation scale exhibit the usual perturbative 
behavior, and no apparent jet quenching should appear. 

\vskip0.3cm

The effects of our suppression
mechanism begin to fail around $6~$ GeV for the $130~$ GeV data and somewhat higher for the $200~$ GeV
data.  We might have been able to work harder and find some marginally acceptable parameterization
of the gluon structure functions which would extend the agreement to larger $p_T$, but this would not
be very natural.  Part of the problem is at these large $p_T$ values, we become sensitive to the
gluon structure function within the nuclear fragmentation region $x > 0.1$, and this is a region
which is more complicated than that for which our naive saturation model can accommodate.  It is also true that
there must be some effects of jet energy loss in the media produced in heavy ion collisions, and the degree
to which our saturation mechanism is dominant or subdominant relative to this mechanism is impossible
for us to assess a priori.  
\vskip0.3cm

The $dA$ data at RHIC will allow to disentangle the effects of jet energy loss from gluon saturation.
Saturation is a property of the nuclear wavefunction, and in so far as it modifies jet production
in $AA$ collisions, there will be a corresponding effect in $dA$ collisions.  
Let us discuss this in more detail. In $dA$ collisions at RHIC energy, with the exception of peripheral 
interactions where the density of participants, and thus the saturation scale, is too small, 
the gluon distribution in the 
nucleus $A$ is saturated and in the domain $Q_s < Q < Q_s^2/\Lambda$ is thus given by 
\beq
\varphi_A \sim {S_A \kappa_A \over \alpha_s} \sim {S_A \over \alpha_s} 
\left({Q_{s,A}^2 \over Q^2}\right)^{1/2}.
\eeq
The deuteron distribution is given by $\varphi_d \sim S_d \kappa_d/\alpha_s$; our conclusion on the 
scaling with the number of participants from the nucleus $A$ will be independent on the explicit 
assumption about $\varphi_d$.    
The dependence of semi--hard processes in the saturation region on the number of participants 
from the nucleus $A$ $N_{part}^A$ will now be given by 
\beq \label{NPARTDA}
\frac{d N^{dA}}{d y d^2 p_t}\,\, = \ \frac{1}{S_A} E {d \sigma^{dA} \over d^3 p} \,\
\sim (Q_{s,A}^2)^{1/2} \sim \left(N_{part}^A \right)^{1/2}.
\eeq
where we have used that in $dA$ collisions $Q^2_{s,A} \sim N_{part}^A$.
We predict the behavior (\ref{NPARTDA}) for semi--hard processes to hold in $dA$ collisions at RHIC with the exception of peripheral collisions, where the saturation 
scale $Q^2_{s,A}$ is too small. The yield of particles per participant from the nucleus $A$ in these collisions is thus predicted to decrease as $(N_{part}^A)^{-1/2}$, deviating from the scaling with the number 
of collisions expected in perturbative QCD. 
We expect the saturation effects to set in around $N_{part}^{Au} \simeq 6$, 
corresponding to the impact parameter of $d Au$ collision $b = 5 \div 6$ fm \cite{KLN1}. We thus predict that around 
$N_{part}^{Au} \simeq 6$ the yields of high $p_t$ particles 
would begin to deviate from the scaling with the number of collisions $N_{coll} \sim N_{part}^{Au}$; the yield per 
participant will start to decrease as $(N_{part}^{Au})^{-1/2}$. In $15 \%$ most central $d Au$ events, 
where $N_{part}^{Au} \simeq 12$ 
\cite{KLN1}, we therefore expect to see the normalized yield of $(6/12)^{1/2}\simeq 0.7$, corresponding 
to $\simeq 30 \%$ suppression of high $p_t$ particles\footnote{Numerical calculations according 
to the formulae given above give somewhat smaller, but close, $\simeq 25 \%$ suppression effect.}. 
The scaling of semi--hard processes in $d Au$ collisions with centrality at RHIC energy can 
thus expected to be drastically different from that observed previously at fixed target energies, 
where at high $p_t$ the yields of particles were proportional to the number of collisions.

On the other hand,
jet energy loss is expected to be a minor effect for $dA$ collisions since the jets are produced
in a region far from the fragments of the nucleus \cite{Wang:1998ww}; 
for recent predictions, see \cite{Vitev:2002pf}. 

\vskip0.3cm

In addition, the $dA$ collisions provide the opportunity
of studying jet production in the fragmentation region of the deuteron, and this probes much smaller
values of $x$ of the nucleus than is the case for centrally produced jets. One expects the saturation 
scale  to grow as \cite{KL}  
$Q_{s,A}^2(y) = Q_{s,A}^2(0) \exp(\lambda y)$, with $\lambda \simeq 0.25$. 
At 3 units of rapidity away from the central rapidity region toward the deuteron fragmentation 
region in $dA$ collisions at $\sqrt{s} = 200$ GeV, the saturation scale thus 
grows to about $Q_{s,A}^2(y=3) \simeq 4.5\ \rm{GeV}^2$. The approximations we
are forced to make for centrally produced jets in $AA$ collisions should be under much better control
for jets arising in the fragmentation region of the deuteron; recently, this problem was addressed 
in \cite{Dumitru:2001ux}, \cite{Dumitru:2002qt},  \cite{Gelis:2002ki}, \cite{Dumitru:2001jn}, \cite{Lenaghan:2002js}, \cite{Gelis:2002nn}.

\end{document}